\documentclass[preprint,showpacs,preprintnumbers,amsmath,amssymb]{revtex4}
\usepackage{graphicx}
\usepackage{dcolumn}
\usepackage{bm}
\def\etal{{\it et al.}}

\def\ubar{\overline{u}}

\def\gam5{\gamma_5} 
\def\gamA{\gamma^{\alpha}}
\def\gama{\gamma_{\alpha}}

\def\sigmaprim2{\sigma^{\prime 2}}
\def\piprim2{\vec{\pi}^{\prime 2}} 

\begin{document} 
\begin{center}
\vspace{5mm}
\Large{\bf Ordinary Muon Capture in Hydrogen Reexamined}
\\
{\large U. Raha}$^{1,2}$, {\large F. Myhrer}$^1$,  
{\large and K. Kubodera}${^1}$\\
{$^1$Department of Physics and Astronomy,\\
University of South Carolina,\\ 
Columbia, SC 29208, USA }\\
{$^2$Department of Physics, Indian Institute of Technology, 
Guwahati-781 039 Assam, India}

\end{center}
\centerline{(\today)}
\vskip 1cm 

\vspace*{10mm}

\noindent
ABSTRACT ---   The rate of muon capture in a muonic hydrogen atom
is calculated in heavy-nucleon chiral perturbation theory
up to next-to-next-to leading order.
To this order, we present the systematic evaluation 
of all the corrections due to the QED 
and electroweak radiative corrections
and the proton-size effect.
Since the low-energy constants involved
can be determined from other independent sources of information,
the theory has predictive power.
For the hyperfine-singlet $\mu p$ capture rate $\Gamma_0$,
our calculation gives
$\Gamma_0=710 \,\pm 5\,s^{-1}$,
which is in excellent agreement with the experimental value
obtained in a recent high-precision measurement
by the MuCap Collaboration.

\newpage

\section{Introduction}

In a recent MuCap Collaboration experiment~\cite{MuCap2007},
the rate $\Gamma_0$ of muon capture from the hyperfine-singlet state
of a $\mu p$ atom was measured to 1 \% precision.
The reported experimental value is
\begin{eqnarray}
\Gamma_0^{\rm exp}(\mu^- p \to \nu_\mu n) 
&=& 
714.9 \pm 5.4 (stat) \pm 5.1 (syst)\, {\rm sec}^{-1}
\, .
\label{eq:Gammaexp}
\end{eqnarray}
As is well known~\cite{Primakoff1975},
the $\mu p$ capture process is the primary source of information
on the pseudoscalar form factor, $G_P(q^2)$,
which appears in the nucleon matrix element of 
the axial-vector weak
current (see Eq.(\ref{eq:N-currents}));
for recent reviews, see~\cite{Fearing2003,Kammel2010}.
To be more specific,
$\mu p$ capture is sensitive to the quantity 
$g_P\equiv G_P(q^2\!=\!-0.88 m_\mu^2)$,
where $q^2$
is the four-momentum transfer squared
relevant to $\mu p$ capture ($m_\mu$ is the muon mass). 
Bernard {\it et al.}~\cite{bkm94} used heavy baryon chiral perturbation theory 
(HB$\chi$PT) to calculate $G_P(q^2)$;
their results essentially reproduce those obtained earlier 
by Adler and Dothan~\cite{Adler1966} based on PCAC,
and by Wolfenstein~\cite{Wolfenstein1970} with the use of 
the dispersion relations.
It is to be emphasized, however, that the systematic 
expansion scheme of HB$\chi$PT allows us to conclude 
that the corrections to the expression for $G_P(q^2)$
obtained by Bernard {\it et al.} are very small~\cite{Kaiser2003}.
The value of $g_P$ derived from HB$\chi$PT is
$g_P=8.26\pm0.23$~\cite{bkm94}.
Meanwhile, the empirical value of $g_P$ extracted 
from $\Gamma_0^{\rm exp}$ with the use of the theoretical framework 
provided in Ref.\cite{Sirlin2007} is  
$g_P^{exp}= \! 8.06 \pm 0.55 $~\cite{MuCap2007},
which is consistent with the theoretical value.

It is to be noted that a theoretical treatment of $\mu p$ capture
that matches the 1~\% experimental accuracy
requires a rigorous treatment of the radiative corrections (RCs)
of order $\alpha$.
Czarnecki {\it et al.}~\cite{Sirlin2007} 
calculated the relevant RCs within the theoretical framework 
developed by Sirlin and Marciano~\cite{Sirlin1974, Sirlin1986}. 
In this approach (to be referred to as the S-M approach),
the RCs of order $\alpha$ are decomposed into so-called
``outer" and ``inner" corrections.
The outer correction is a universal function of the lepton velocity
and is model-independent, whereas the inner correction
is affected by the short-distance physics and hadron structure.
The inner corrections arising from photon and weak-boson loop
diagrams are divided into high-momentum and low-momentum contributions.
The former is evaluated in the current-quark picture, 
while the latter is estimated with the use 
of the phenomenological electroweak nucleon form factors.
The expression for $\Gamma_0$ including the RCs of order $\alpha$
due to Czarnecki {\it et al.}~\cite{Sirlin2007} 
was used by the MuCap Group
in deducing the above-mentioned value of $g_P^{exp}$ 
from $\Gamma_0^{\rm exp}$.
We remark that, although the estimates of inner corrections
in the S-M approach are considered to be reliable to the level of accuracy
quoted in the literature, the possibility that these estimates
may contain some degree of model dependence 
is not totally excluded.  
This motivates us to present here 
a calculation of the RCs of order $\alpha$ based on 
model-independent effective field theory (EFT).

In this note we evaluate the RCs for $\mu p$ capture 
based on HB$\chi$PT, 
which is an effective low-energy field theory of QCD, 
see e.g., Refs.~\cite{bkm95,Bernard2009,Scherer}. 
In HB$\chi$PT, the short-distance hadronic and electroweak processes
are subsumed into a well-defined set of low-energy constants (LECs),
which means that the LECs should systematically parametrize 
the inner corrections. 
Therefore, provided that there are sufficient sources of information 
to fix these LECs, the HB$\chi$PT approach gives model-independent results
with the possibility to  
estimate higher-order corrections. 
In two of the earlier publications we used the same EFT approach
to evaluate RCs to order $\alpha$ 
for the neutron $\beta$-decay~\cite{Ando2004}, and  
for the  inverse $\beta$-decay reaction, $\bar{\nu}_e  p \to e^+  n$,
at low energies~\cite{Udit2011}. 
It is to be noted that the EFT treatments 
of the $\mu p$ capture process, neutron $\beta$-decay and 
the $\bar{\nu}_e p \to e^+  n$ reaction involve 
the same LECs. 
Therefore, if we can determine these LECs 
with the use of experimental information for one process,
we can make model-independent predictions for observables 
for the other reactions.

The remainder of this article is organized as follows.
In Section II we explain the basic ingredients that enter 
into the HB$\chi$PT calculation of the $\mu p$ capture rate.  
We describe in Section III the evaluation of the RCs to order $\alpha$,
and give in Section IV the numerical results for the $\mu p$ capture rate
including the RCs. 
Finally, Section V is dedicated to discussion and conclusions.

\section{HB$\chi$PT calculation of the $\mu p$ capture rate}

%
The theoretical framework is essentially  the same as the one 
employed in Ref.~\cite{Ando2004}. 
We therefore present here only 
a brief recapitulation of our formalism, 
relegating details to Ref.~\cite{Ando2004}.
HB$\chi$PT assumes that the characteristic four-momentum for the process, 
$Q \ll \Lambda_\chi  \simeq 1$ GeV, 
where  $\Lambda_\chi$ is the 
chiral scale. This theory contains two perturbative expansions, 
one in terms of the expansion parameter $Q/\Lambda_\chi \ll 1$ 
and the other in terms of  $Q/m_N \ll 1$, where $m_N$ is the nucleon mass. 
Since $m_N \simeq \Lambda_\chi$,
the two expansions are considered simultaneously. 
%
%
When we include RCs in our considerations,  
a third expansion parameter $\alpha$ enters the theory. 
Our concern here is to carry out a HB$\chi$PT calculation
up to next-to-next-to-leading order (NNLO), 
i.e., to order $(Q/\Lambda_\chi)^2 \simeq \alpha\simeq 1/137$.
In what follows, we first describe the contributions
that arise from the expansions in $Q/\Lambda_\chi$ and $Q/m_N$.
This part is based on the previous HB$\chi$PT results 
that can be found in Refs.~\cite{Ando2000,bhm2001,Bernard2002}. 
(We follow the notations used in Ref.~\cite{Ando2000}.)
We then proceed to explain our calculation of RCs.

Muon capture being a low-energy process,
the relevant weak interaction can be expressed as 
the local current-current interaction, and 
the transition amplitude for the ordinary muon capture 
(OMC) process in hydrogen, $\mu^-p\rightarrow\nu_{\mu}n$\,, 
is given by
\begin{eqnarray}
\label{mfi}
{\mathcal M}_{fi}\!&=&\!\frac{G_\beta}{\sqrt{2}}
\langle n\nu_{\mu}|\hat{l}_{\alpha}\hat{j}^{\alpha}|(\mu^- p)_{\rm atom}
\rangle
\approx 
\frac{G_{\beta}}{\sqrt{2}}
\sqrt{\frac{m_\mu\!+\!m_N}{2m_\mu  m_N}}\Psi_{\mu p}({\mathbf 0})
\langle n\nu_\mu |\hat{l}_0\hat{j}^0
\!\!-\!\hat{{\mathbf l}}\!\cdot\!\hat{{\mathbf j}}
|\mu p\rangle
%
\nonumber\\ &\equiv&
\frac{{\mathcal N}_{\rm rel}G_{\beta} }{2}
\sqrt{\frac{m_\mu\!+\!m_N}{2m_{\mu}m_N}}
\Psi_{\mu p}({\mathbf 0}){\mathcal T}_{\rm NR}\,. 
\end{eqnarray} 
In the above, $G_\beta\equiv G_F V_{ud}$,   
where $G_F$ is the Fermi coupling constant determined 
from the muon decay rate,
and the $V_{ud}$ is the CKM matrix element given in Ref.~\cite{RPP2012}.
$\Psi_{\mu p}({\mathbf 0})$ is the $\mu p$-atomic wave function 
at ${\mathbf r} \!=\! 0$. 
The normalization factor ${\mathcal N}_{\rm rel}$,
which arises from ``matching'' between the standard relativistic normalization
of spinors and the corresponding non-relativistic normalizations
is given by 
${\mathcal N}_{\rm rel}=
4m_N\!\sqrt{m_\mu E_\nu}$\,,
where 
\begin{eqnarray}
E_\nu=\frac{(m_{\mu}\!+\!m_p)^2\!-\!m^2_n}{2(m_{\mu}\!+\!m_p)}
=99.149\, {\rm MeV}\, .
\end{eqnarray}
The non-relativistic transition amplitude, ${\mathcal T}_{\rm NR}$, 
in Eq.(\ref{mfi}) is written as
\begin{eqnarray}
{\mathcal T}_{\rm NR}=\chi^{\dagger}_n\chi^{\dagger}_\nu\,
\widehat{\mathcal M}\,\chi_\mu\chi_p\,, 
\label{eq:tfi}
\end{eqnarray}
where $\chi_{p,n}$ and $\chi_{\mu,\nu}$ are the nucleon and lepton two-spinors, 
respectively; the explicit expression for the operator
$\widehat{\mathcal M}$ 
will be given in what follows.

The matrix element of the leptonic weak current operator, $\hat{l}_{\alpha}$
in Eq.(\ref{mfi}) is given by
$l_{\alpha}$ $\equiv\langle\nu|\hat{l}_\alpha|\mu\rangle=$
$\ubar_{\nu}\gama(1\!-\!\gam5) u_{\mu}$\,,
which in the present case takes the form 
\begin{equation}
l_0 = \frac{1}{\sqrt{2}}\,
\chi^{\dagger}_\nu(1\!-\!\vec{\sigma}\cdot\hat{\nu})\chi_\mu\,,
\,\,\,\,\,\,\, 
{\mathbf l} = \frac{-1}{\sqrt{2}}\,
\chi^{\dagger}_\nu(1\!-\!\vec{\sigma}\cdot\hat{\nu})
\vec{\sigma}\chi_\mu\,, 
\end{equation} 
where $\hat{\nu}$ is the unit vector 
in the  direction of the neutrino momentum. 
The matrix elements of the nucleon weak current operator,
$\hat{j}^{\alpha}\!=\hat{j}^{\alpha}_v\!-\!\hat{j}^{\alpha}_a$\,,
are given by\footnote{We assume here the absence of second class currents.} 
\begin{eqnarray}
\langle n(p')|\hat{j}^\alpha_v|p(p)\rangle &\equiv&
j^{\alpha}_v \,=\,\ubar_n(p')\!\left[F^v_1(q^2)\gamA \!+\! 
F^v_2(q^2)\frac{i\sigma^{\alpha\beta}q_\beta}{2m_N}\right]\!u_p(p)
\nonumber\\
\langle n(p')|\hat{j}^\alpha_a|p(p)\rangle &\equiv&
j^{\alpha}_a\,=\,\ubar_n(p')\!\left[G_A(q^2)\gamA\!\gam5\! +\! 
G_P(q^2)\frac{q_\beta}{m_\mu}\gam5\right]\!u_p(p)\,, 
\label{eq:N-currents}
\end{eqnarray}
where $F^v_1(q^2)$, $F^v_2(q^2)$, $G_A(q^2)$ and $G_P(q^2)$ are the vector, 
weak-magnetism, axial-vector and pseudoscalar form factors, respectively, 
and where the  $m_N$ is the average nucleon mass,
$m_N\!=\!\frac{1}{2}(m_p+m_n)$. 
In the {\it rest frame of the initial proton},
the non-relativistic nucleon currents 
in HB$\chi$PT  are given by\footnote{ 
We utilize the heavy-mass decompositions: 
$p'_{\mu}\rightarrow m_Nv_{\mu}+ r'_{\mu}$ and $p_{\mu}\rightarrow m_Nv_{\mu}$. 
}
\begin{eqnarray}
j^\alpha_v\!&=&\!{\mathcal N}_n\,{\bar n}_v(p')
  \!\left\{\left[\frac{2m_N}{E'\!+\!m_N}F^v_1(q^2)- 
\frac{E'\!-\!m_N}{E'\!+\!m_N}F^v_2(q^2)\right]v_\alpha 
\right.\nonumber \\
&&\left.\times\left[\frac{1}{E'\!+\!m_N}(F^v_1(q^2)\!+\!F^v_2(q^2))-
\frac{1}{2m_N}F^v_2(q^2)\right]q_\alpha 
\right.\nonumber \\
&&\left.+\,\frac{2}{E'\!+\!m_N}
[S_\alpha,S\!\cdot\! q\,](F^v_1(q^2)\!+\!F^v_2(q^2))\right\}p_v(0)\,, 
\nonumber
\\
j^\alpha_a\!\!&=&\!\!{\mathcal N}_n\,{\bar n}_v(p')\!
 \left\{G_A(q^2)\!
 \left[2S_\alpha\!-\!\frac{2(S\!\cdot\!q) \,v_\alpha}{E'\!+\!m_N}\right]
\right.\nonumber
\\
&&\left.\;\;\;\;\;\;\;\;\;\;\;\;\;\;\;\;\;\;\;\;\;\;\;\;\;\;\;\;\;\;\;\;+\,G_P(q^2)
\frac{2(S\!\cdot \!q) \,q_\alpha}{m_\mu (E'\!+\!m_N)}\right\}p_v(0)\,,
\label{eq:NR-currents}
\end{eqnarray}
with the heavy nucleon spinors, $n_v(r')$ and $p_v(0)$
defined as~\cite{bkm95}
\begin{eqnarray}
n_v(p')=\sqrt{\frac{2m_N}{E'\!+\!m_N }}\, \frac{1}{2} (1\!+\!v\!\!\!/)u_n(p')\,,\quad p_v(0)
=\frac{1}{2}(1\!+\!v\!\!\!/)u_p(p)\, . 
\end{eqnarray} 
%
%
The kinematics in the rest-frame of the proton is as follows.
The four-momenta of the initial proton and 
the outgoing neutron are $p=\!(m_N,{\bf 0})$ 
and $p'=\!(E',{\bf p}')$, respectively, where 
$E'\!\!=\!\sqrt{m^2_N\!+\!{\bf p}'^2}$ and ${\bf p}'\!=\!-{\bf p}_{\nu}$. 
The four-momentum transfer in the OMC process 
is $q=p'\!-\!p\!=(q_0,{\bf q})$, 
with
$q_0\!=E'\!-\!m_N\!=
\frac{{\bf p}_{\nu}^2}{2m_N}+\!{\mathcal O}(m^{-2}_N)$ 
and ${\bf q}=\!-{\bf p}_\nu$\,.
Expanding the proton and neutron spinors in Eq.(\ref{eq:NR-currents})
up to ${\mathcal O}(m^{-2}_N)$ leads to
\begin{eqnarray}
j_0(q)\!\!&=&\!\!{\chi}_{n}^{\dagger}\left[f^v_1(q)+
(\vec{\sigma}\cdot\hat{\nu})f^a_3(q)\right]\chi_p\,,
\nonumber\\
{\mathbf j}(q)\!\!&=&\!\!-{\chi}_{n}^{\dagger}
\left[i(\vec{\sigma}\times\hat{\nu})f^v_2(q)+
\hat{\nu}f^v_3(q)+
\vec{\sigma}f^a_1(q)+\hat{\nu}(\vec{\sigma}\cdot\hat{\nu})f^a_2(q)\right]\chi_p\,,
\end{eqnarray}
where the non-relativistic polar-vector form factors are related to the 
standard Lorentz covariant form factors in the proton rest frame via 
\begin{eqnarray*}
f^v_1(q)=F^v_1(q^2)\left(1\!-\!\frac{q^2}{8m^2_N}\right)+\frac{q^2}{4m^2_N}\,F^v_2(q^2)\,,
\end{eqnarray*}
\begin{eqnarray}
f^v_2(q) = \frac{|\mathbf{q}|}{2m_N}\left[F^v_1(q^2)\!+\!F^v_2(q^2)\right]\,,
\quad f^v_3(q)= \frac{|\mathbf{q}|}{2m_N}F^v_1(q^2)\,,
\label{eq:ff1}
\end{eqnarray}
while the non-relativistic axial-vector form factors are related to the 
covariant axial form factors via
\begin{eqnarray}
f^a_1(q)&=&G_{\!A}(q^2)\!\left(\!1-\!\frac{q^2}{8m^2_N}\right)\,, \quad
f^a_2(q) = -\frac{|\mathbf{q}|^2}{2m_{\mu}m_N}
\left(\!1+\frac{q^2}{8m^2_N}\right)\!G_{\!P}(q^2)\,,
\label{eq:ff2}
\end{eqnarray}
\begin{eqnarray}
f^a_3(q)=\frac{|\mathbf{q}|}{2m_N}
\left(\!G_{\!A}(q^2)+
\frac{q^2}{2m_\mu m_N}\,G_{\!P}(q^2)\right).
\label{eq:ff3}
\end{eqnarray}
The non-relativistic form factors 
appearing in Eqs.~(\ref{eq:ff1}), (\ref{eq:ff2}) and (\ref{eq:ff3}) 
have been calculated 
in Refs.~\cite{Ando2000,bhm2001,Bernard1998,Fearing1997},
up to next-to-next-to leading order (NNLO) 
or ${\mathcal O}((Q/\Lambda_\chi)^3)$, in HB$\chi$PT.
In the proton rest-frame, they are given by 
\begin{eqnarray}
f_1^v(q) \!\!&=&\!\! 1+\kappa_V\frac{q^2}{4m^2_N}
\nonumber\\
&&-\,\frac{1}{(4\pi f_{\pi})^2}\left\{q^2\!\left(\frac{2}{3}\,g^2_A+2B^{(r)}_{10}\right)
\!+q^2\!\left(\frac{5}{3}\,g^2_A+\frac{1}{3}\right)\ln\!\left[\frac{M_{\pi}}{\lambda}\right]
\right.\nonumber\\
&&\left.-\int^{1}_{0}\!\!dz
\Big[ M^2_{\pi}(3g^2_A\!+\!1)-q^2z(1\!-\!z)(5g^2_A\!+\!1)\Big]
\ln\!\left[1-z(1\!-\!z)\frac{q^2}{M^2_\pi}\right]\right\}\,,
\nonumber\\
f_2^v(q) \!\!&=&\!\! \frac{|\mathbf{q}|}{2m_N} 
\left\{ 1+\kappa_V+g^2_A
\frac{4\pi m_NM_\pi}{(4\pi  f_{\pi})^2}\!\int^1_0\!\!dz\!
 \left[1-\!\sqrt{1-\!z(1\!-\!z)\frac{q^2}{M^2_\pi}}\, \right] \right\} \,,
\nonumber\\
f_3^v(q) \!\!&=&\!\! \frac{|\mathbf{q}|}{2m_N}\,,
\nonumber\\
f_1^a(q) \!\!&=&\!\! g_A\!\left(\!1\!-\!\frac{q^2}{8m^2_N}\right) +
\frac{q^2}{(4\pi f_{\pi})^2}{\tilde{B}}_3\,,
\nonumber\\
f_2^a(q) \!\!&=&\!\!
\frac{|\mathbf{q}|^2}{q^2\!-\!M^2_\pi}
\left\{g_A\!\left(\!1\!+\!\frac{q^2}{8m^2_N}\right)\!-
    \frac{2M^2_\pi}{(4\pi f_{\pi})^2}{\tilde{B}}_2\right\}+
    \frac{|\mathbf{q}|^2}{(4\pi f_{\pi})^2}\tilde{B}_3\,,
\nonumber\\
f_3^v(q) \!\!&=&\!\! \frac{|\mathbf{q}|}{2m_N}g_A\!
       \left(\!1\!-\!\frac{q^2}{q^2\!-\!M^2_\pi}\right)\,. 
\label{eq:ff}
\end{eqnarray}

\vspace{5mm}
\noindent
In terms of the quantities derived above,
the operator $\widehat{\mathcal M}$ [see Eq.(\ref{mfi})]
is written as
\begin{eqnarray}
\widehat{\mathcal M} &=& \left(1\!-\!\vec{\sigma}_l\!\cdot\!\hat{\nu}\right)
\left[\,f^v_1(q)-
i\vec{\sigma}_l\!\cdot\!(\vec{\sigma}_N\!\times\!\hat{\nu})f^v_2(q)-
(\vec{\sigma}_l\!\cdot\!\hat{\nu})f^v_3(q)
\right. 
\nonumber \\ && \left. 
- (\vec{\sigma}_l\!\cdot\!\vec{\sigma}_N)f^a_1(q)-
(\vec{\sigma}_l\cdot\hat{\nu})(\vec{\sigma}_N\!\cdot\!\hat{\nu})f^a_2(q)+
(\vec{\sigma}_N\!\cdot\!\hat{\nu})f^a_3(q)\,\right]\,,
\end{eqnarray}
where $\vec{\sigma}_l$ and $\vec{\sigma}_N$
are the spin matrices acting on the lepton and nucleon spinors,
respectively.
We may choose the direction of the emitted neutrino 
as our $z$-axis, i.e., $\hat{\nu}\equiv\hat{z}$. 
In the helicity basis, the amplitude
${\mathcal T}_{\rm NR}$ appearing in Eqs.(\ref{mfi}) and (\ref{eq:tfi}) 
is given as
\begin{eqnarray}
{\mathcal T}_{\rm NR}
\equiv\widetilde{M}(h;S,S_z)
=\sum_{S^p_z,S^\mu_z}
\langle\frac{1}{2}S^p_z;\frac{1}{2}S^\mu_z\mid\frac{1}{2}\frac{1}{2};SS_z\rangle\,
\langle\frac{1}{2}S^n_z;\frac{1}{2}, \frac{-1}{2}|\widehat{\mathcal M}|
\frac{1}{2}S^p_z;\frac{1}{2}S^\mu_z\rangle 
\label{eq:Mtilde}
\end{eqnarray}
where $h\!=\!L$ ($h\!=\!R$) corresponds
to the positive (negative) helicity state of the final-state neutron,
and $S\!=\!0$ ($S\!=\!1$) represents the hyperfine-singlet (triplet)
state of the muonic hydrogen atom \footnote{
The relation between $\widetilde{M}(h;S,S_z)$
and the helicity amplitude $M(h;S,S_z)$ used by Ando \etal~\cite{Ando2000} 
is as follows:
$M(+;S,S_z)=\frac{{\mathcal N}_{\rm rel}G_{\beta}}{2}
\widetilde{M}(L;S,S_z)$,
$M(-;S,S_z)=\frac{{\mathcal N}_{\rm rel}G_{\beta}}{2}
\widetilde{M}(R;S,S_z)$.}.
The constraint,
$S_z=S^n_z\!-\!\frac{1}{2}$, reduces the eight possible
helicity amplitudes in Eq.(\ref{eq:Mtilde})  
to the following three: 
%
\begin{eqnarray}
\widetilde{M}(L;0,0)&=&
\sqrt{2}\left(f^v_1+2f^v_2+f^v_3+3f^a_1+f^a_2+f^a_3\right)\,,
\nonumber\\
\widetilde{M}(L;1,0)&=&
\sqrt{2}\left(f^v_1-2f^v_2+f^v_3-f^a_1+f^a_2+f^a_3\right)\,,
\nonumber\\
\widetilde{M}(R;1,-1)&=&
2\left(f^v_1+f^v_3-f^a_1-f^a_2-f^a_3\right)\,. 
\label{eq:hel-amp}
\end{eqnarray}
Finally, since the binding energy of the
muonic hydrogen atom can safely be ignored, 
the total OMC rate in a hyperfine state $S$ is given as
%
\begin{eqnarray}
\Gamma_S & \!= \!& \frac{1}{2(m_\mu\!+\!m_N)}\!\cdot\!\frac{1}{2S\!+\!1}
       \int\!\frac{d^3{\mathbf p}'}{(2\pi)^32E'}\frac{d^3{\mathbf p}_{\nu}}{(2\pi)^32E_{\nu}}\,
        (2\pi)^4\delta^4(P_I-p'-p_{\nu})\left|{\mathcal M}_{fi}\right|^2
\nonumber\\
&\! =\! &\frac{G_{\beta}^2{\mathcal N}_{\rm rel}^2}
{16m_\mu m_N}\!\cdot\!\frac{1}{2S\!+\!1}
\left|\Psi_{\mu p}({\mathbf 0})\right|^2 
\!\!\int\!\!\frac{d^3{\mathbf p}'}{(2\pi)^32E'}\frac{d^3{\mathbf p}_{\nu}}{(2\pi)^32E_{\nu}}\,
 (2\pi)^4\delta^4(P_I\!-\!p'\!-\!p_{\nu})\sum_{S_z,h}\left|\widetilde{M}(h;S,S_z)\right|^2
\nonumber\\
&\!=\! & \frac{G_{\beta}^2{\mathcal N}_{\rm rel}^2}{2S\!+\!1}\!\cdot\!
\frac{\left|\Psi_{\mu p}({\mathbf  0})\right|^2}{64\pi\,m_\mu  m_N}
\left(\frac{E_\nu}{E_\nu\!+\!\!\sqrt{m^2_N\!+\!E^2_\nu}}\right)
\sum_{S_z,h}\left|\widetilde{M}(h;S,S_z)\right|^2
\, , 
\label{eq:C-rate}
\end{eqnarray} 
where $P_I$ is the initial total four-momentum. 
%
%
If we ignore radiative corrections and identify 
$\Psi_{\mu p}({\mathbf 0})$
with the lowest-order $1s$-state Coulomb wave function, 
$\Phi_{1s}({\mathbf  0})=(\alpha^3\mu^3_{\mu p}/\pi)^{\!1/2}$
with $\mu_{\mu p}\! \equiv\! m_\mu m_p / (m_\mu\! + \!m_p)$,
then the last line in Eq.(\ref{eq:C-rate})
agrees with Eq.(26) in Ref.~\cite{Ando2000}.
In the present work, however, we do include radiative corrections,
and it turns out 
that, at ${\cal O}(\alpha )$ under consideration,
there appear two types of significant radiative corrections 
to $\Phi_{1s}({\mathbf  0})$,
and these corrections will be discussed in the following section.

\vspace{3mm} 
We now evaluate the form factors in Eq.(\ref{eq:ff}), which determine 
the helicity amplitudes in Eq.(\ref{eq:hel-amp}).
{}Table~\ref{tab:ff} shows the numerical values of
the nucleon weak form factors and helicity amplitudes
calculated for the four-momentum transfer,
$q^2=q^2_*\equiv-0.88m^2_\mu$, relevant to OMC.
These numerical values were obtained with the use of 
the following input parameters:
$g_A=1.266$, $\kappa_V = 3.706$, $f_\pi=92.42$ MeV, $M_\pi=139.57$ MeV, 
and $m_N= 938.919$ MeV. 
The LECs appearing in Eq.(\ref{eq:ff}) are determined following 
Refs.~\cite{bkm95,Bernard1998}. 
First, ${\tilde{B}}_2$ is fixed from the 
Goldberger-Treiman (G-T) discrepancy relation, 
\begin{eqnarray}
\frac{2M_\pi^2}{(4\pi f_\pi)^2\, g_A}\, {\tilde{B}}_2 = 
\frac{g_A\, m_N}{g_{\pi NN} f_\pi} -1\, .
\label{eq:B2-GT}
\end{eqnarray} 
For $g_{\pi NN}\!=\!13.40$ and $g_A\!=\!1.266$
(see, e.g., PDG2002~\cite{PDG2002}),
this relation leads to ${\tilde{B}}_2= -1.761$. 
The values of $g_A$ and $g_{\pi NN}$
have been slightly changing over the years; 
if we use the latest values $g_{\pi NN}\!=\!13.05$ and $g_A\!=\!1.270$
(taken from PDG2012~\cite{RPP2012}),
we obtain ${\tilde{B}}_2=-0.498$.
To what extent 
the existing uncertainties in $g_A$ and $g_{\pi NN}$
affect the calculated $\mu p$ capture rate
will be discussed in the last section. 
The LEC, ${\tilde{B}}_3$,
is fixed from the nucleon axial radius,  
\begin{eqnarray}
{\tilde{B}}_3 = \frac{g_A}{2}\, (4\pi f_\pi)^2 
\, \frac{ \langle r_A^2 \rangle  }{3}\, .
\nonumber
\end{eqnarray} 
The value of the iso-vector axial radius $\langle r_A^2\rangle^{1/2} $
has large uncertainty, see e.g., Ref.~\cite{bkm95}. 
From the empirical axial form factor $G_A(t)= g_A/(1\!-\!t/m_A^2)^2$, we find
$\langle r_A^2 \rangle^{1/2} = \sqrt{12}/m_A$ = 0.62 fm (0.57 fm) for 
$m_A \!=\! 1100$ MeV (1200 MeV). 
We adopt the value $\langle r_A^2 \rangle^{1/2} = 0.65$ fm cited in 
Ref.~\cite{Bernard1998} to find ${\tilde{B}}_3 = 3.08 $. 
The last of the LECs in Eq.(\ref{eq:ff}), ${\tilde{B}}_{10}^{(r)}$,
is related to the nucleon iso-vector form factor~\cite{Bernard1998}  
\begin{eqnarray}
\frac{1}{6} \,\langle r_V^2 \rangle = 
-\frac{2 {\tilde{B}}_{10}^{(r)}\!(\Lambda_\chi ) }{(4\pi f_\pi)^2} - 
\frac{1\!+\!7g_A^2}{6 (4\pi f_\pi)^2} - 
\frac{5g_A^2\!+\!1}{3(4\pi f_\pi)^2 }\,
{\rm ln}\!\left(\!\frac{M_\pi}{\Lambda_\chi}\!\right)\,.
\nonumber
\end{eqnarray}
From the measured value of $\langle r_V^2 \rangle$
we obtain ${\tilde{B}}_{10}^{(r)} = 0.27$
for  $\Lambda_\chi = 1$ GeV. 
These values of the LECs were used 
in obtaining the numerical results given 
in Table~\ref{tab:ff}. 
\begin{table}[h]
\caption{The OMC form factors Eq.(\ref{eq:ff}) and 
helicity amplitudes Eq.(\ref{eq:hel-amp}) 
at $q^2_*=-0.88m^2_\mu$, 
obtained for $g_A = 1.266$ and $g_{\pi NN}=13.40$.}

\vspace{3mm}

\begin{tabular}{c|c|c|c|c|c|c|c|c}
\hline\hline
$f^v_1(q_*)$ \, & \, $f^v_2(q_*)$ \, & \, $f^v_3(q_*)$ \, & \, $f^a_1(q_*)$ \,
& \, $f^a_2(q_*)$ \, & \, $f^a_3(q_*)$ \, &  $\widetilde{M}(L;0,0)$  &
 $\widetilde{M}(L;1,0)$  &  $\widetilde{M}(L;1,-1)$ \\
\hline
 0.966 \, & \, 0.244 \, &  \, 0.053 \, & \, 1.245 \, &
 \, -0.419 \, & \, 0.044 \, &  3.447   &  -0.770   &  0.148 \\
\hline
\end{tabular} 
\label{tab:ff} 
\end{table}

\section{Radiative corrections}
In this section we consider radiative corrections to OMC,
which  consist of the usual QED loop corrections 
and loop corrections involving a weak-interaction vertex.
Relegating the discussion of the latter 
to the end of the section, we first discuss the QED loop corrections.

The initial state in $\mu^- p$ capture is a charge-neutral $\mu^- p$-atom,
and the final state involves only electrically neutral particles.
Therefore, to the order in HB$\chi$PT under consideration, 
the bremsstrahlung process does not contribute 
to the ``standard" radiative corrections (we ignore the higher order, i.e.,
${\cal O}(\alpha/m_N)$ corrections which are negligible.) 
There are, however, two QED loop corrections to the initial state wave function
which must be considered: the  vacuum polarization correction, 
$\delta\psi^{\rm VP}_{1s}({\mathbf 0})$,
and the correction due to the finite proton size,
$\delta\psi^{\rm FS}_{1s}({\mathbf 0})$.
Inclusion of these corrections 
changes the lowest-order muonic atomic wave function,
$\Phi_{1s}({\mathbf  0})$, into $\Psi_{1s}({\mathbf 0})$:
\begin{eqnarray}
\Psi_{1s}({\mathbf 0}) =\Phi_{1s}({\mathbf  0}) \!
\left[1+\delta\psi^{\rm VP}_{1s}({\mathbf 0})
+\delta\psi^{\rm FS}_{1s}({\mathbf 0})\right].
\label{eq:wavefunction}
\end{eqnarray}
Eiras and Soto~\cite{Eiras2000} calculated 
$\delta\psi^{\rm VP}_{1s}({\mathbf 0})$ to order ${\mathcal O}(\alpha)$, 
while Friar~\cite{Friar1979} discussed 
${\mathcal O}(\alpha)$ contributions to 
$\delta\psi^{\rm FS}_{1s}({\mathbf 0})$.

The analytic expression for 
$\delta\psi^{\rm VP}_{1s}({\mathbf 0})$ derived
by Eiras and Soto~\cite{Eiras2000}  reads
\begin{eqnarray}
\delta\psi^{\rm VP}_{1s}({\mathbf 0})&\!\!\! =\! \! & \! \!
-\frac{\alpha}{\pi}
\left[\left\{\frac{5}{9}-\frac{\pi}{4}\xi+\frac{1}{3}\xi^2-\frac{\pi}{6}\xi^3+
\frac{1}{3}(\xi^4\!+\!\xi^2\!-\!2)F_1(\xi)\right\}_{\gamma\!\!\!/}\right.
\nonumber\\
&+&\!\! \!\left.\left\{\frac{11}{18}-\frac{2}{3}\xi^2+
\frac{2\pi}{3}\xi^3-\frac{12\xi^4\!+\!\xi^2\!+\!2}{6}F_1(\xi)-
\frac{4\xi^4\!+\!\xi^2\!-\!2}{6(\xi^2\!-\!1)}\,
[1\!-\xi^2F_1(\xi)]\right\}_{\rm pole}
\right.
\nonumber\\
&+&\!\!\!\left.\left\{\frac{2}{3}+\frac{\pi}{4}\xi-
\frac{1}{9}\xi^2+\frac{13\pi}{18}\xi^3-
\frac{1}{9}(13\xi^4\!-11\!\xi^2\!-\!11)F_1(\xi)
\right.\right.
\nonumber\\
&&\left.\left.\!\!\!\!\!\!\!\!\!-\frac{1}{3}(4\xi^3\!+\!3\xi)F_2(\xi)\!+\!
\frac{1}{3}(4\xi^4\!+\!\xi^2\!-\!2)F_3(\xi)
\!+\!\frac{1}{3}\!\left(\!4\xi^2\!+\!\frac{11}{3}\!\right)\!
\ln\frac{\xi}{2}\right\}_{\!{\rm multi-}\gamma}\right]
\label{eq:vacuum}
\end{eqnarray}
where 
$\xi\equiv m_e/(\alpha  \mu_{\mu p}) \!\sim\! {\cal O}(1)$;
the expressions for the functions $F_i(\xi)$ ($i=1,2,3$) 
in Eq.(\ref{eq:vacuum}) can be found in Ref.~\cite{Eiras2000}. 
As explained in Ref.~\cite{Eiras2000},  
the first curly bracket in Eq.~(\ref{eq:vacuum})  
corresponds to zero photon exchange contributions, 
the second bracket corresponds to Coulomb pole subtraction terms, 
and the third bracket represents the multi-photon exchange contributions.  
Thus $\delta\psi^{\rm VP}_{1s}({\mathbf 0})$ 
consists of three parts:
\begin{eqnarray}
\delta\psi^{\rm VP}_{1s}({\mathbf 0})=
[\delta\psi^{\rm VP}_{1s}({\mathbf 0})]_{\gamma\!\!\!/}+
[\delta\psi^{\rm VP}_{1s}({\mathbf 0})]_{\rm pole}+
[\delta\psi^{\rm VP}_{1s}({\mathbf 0})]_{{\rm multi}-\gamma}.
\label{eq:threeterms}
\end{eqnarray}
We denote by $\Gamma_S^{(0)}$ 
the $\mu p$ capture rate
for the hyperfine-state $S$ ($S$ = 0 or 1),
obtained by using $\Phi_{1s}({\mathbf  0})$
for $\Psi_{\mu p}({\mathbf 0})$
in Eq.(\ref{eq:C-rate}). 
The use of 
$\Phi_{1s}({\mathbf  0}) \!
\left[1\!+\!\delta\psi^{\rm VP}_{1s}({\mathbf 0})\right]$
for $\Psi_{\mu p}({\mathbf 0})$
in Eq.(\ref{eq:C-rate}) changes $\Gamma_S^{(0)}$ into
\begin{equation}
\Gamma_S^{(0)}\!+ \delta\Gamma_{S}^{\rm VP}
\equiv 
\Gamma_S^{(0)}[1\!+2\delta\psi^{\rm VP}_{1s}({\mathbf 0})].
\end{equation}
Table~\ref{tab:QED} shows the numerical results for
$(\delta\Gamma_{S})^{\!\rm VP}\!/\Gamma_S^{(0)}=2\delta\psi^{\rm VP}_{1s}$.
The first three columns show the individual contributions
of the three terms in Eq.(\ref{eq:threeterms}), while
the fourth column gives 
$2\delta\psi^{\rm VP}_{1s}$, 
which is the sum of these three contributions.
For comparison, in the fifth and sixth columns,
we quote the values of $2\delta\psi^{\rm VP}_{1s}$ (in our notation) 
obtained in Refs.~\cite{Sirlin2007,Goldman1972}. 
\begin{table}[h]
\caption{ Corrections from vacuum polarization (VP) effects, 
          $(\delta\Gamma_{S})^{\!\rm VP}\!/\Gamma_{S}^{(0)}
          =2\delta\psi^{\rm VP}_{1s}({\mathbf 0})$.
          The last two columns give the values of
          $2\delta\psi^{\rm VP}_{1s}({\mathbf 0})$
          in Refs.~\cite{Sirlin2007,Goldman1972} for comparison.}
          
\begin{tabular}{c|c|c|c|c|c}
\hline\hline
 Zero photon & Coulomb pole & Multi-photon & Total VP & {\it Czarnecki,}
 &  \\
  exchange   &   subtraction & exchange   &  contribution & {\it  Marciano}, &
  {\it Goldman} \\
$2[\delta\psi^{\rm VP}_{1s}({\mathbf 0})]_{\gamma\!\!\!/}$ &
$2[\delta\psi^{\rm VP}_{1s}({\mathbf 0})]_{\rm pole}$ &
$2[\delta\psi^{\rm VP}_{1s}({\mathbf 0})]_{{\rm multi}-\gamma}$ &
$ 2 \, \delta\psi^{\rm VP}_{1s}({\mathbf 0})$ & \& {\it Sirlin} \cite{Sirlin2007} 
& \cite{Goldman1972} 
\\[1ex]
\hline
$1.045\frac{\alpha}{\pi}$ & $0.358\frac{\alpha}{\pi}$ &
$0.250\frac{\alpha}{\pi}$ & $1.654\frac{\alpha}{\pi}$ &
$1.73\frac{\alpha}{\pi}$  & $2.95\frac{\alpha}{\pi}$ \\
\hline
\end{tabular}
\label{tab:QED}
\vspace{8mm}
\end{table}
Our result for $ 2 \, \delta\psi^{\rm VP}_{1s}({\mathbf 0})$
agrees with the value given by Czarnecki {\it et al.}~\cite{Sirlin2007} 
within $\sim$5\%. 
Since the size of the $2\delta\psi^{\rm VP}_{1s}$ correction itself
is about 0.4\%, we can say this part of QED corrections
is controlled with sufficient accuracy for our purpose. 

\vspace{2mm}
The proton finite-size correction up to  ${\mathcal O}(\alpha)$
is given as~\cite{Friar1979}
\begin{eqnarray} 
\delta\psi^{\rm FS}_{1s}({\mathbf 0}) 
=-\alpha\mu_{\mu p} \langle r \rangle_p\,.
\label{eq:FS}
\end{eqnarray}
where $\langle r \rangle _p$ is the first moment of the 
proton charge distribution, $\rho_p(\mathbf{r})$.  
Unfortunately, $\langle r \rangle _p$ cannot be measured directly, 
whereas the second moment, $\langle r^2 \rangle _p$,
can be extracted from experimental data. 
In order to evaluate $\langle r \rangle _p$, 
we assume a certain functional form of the proton charge distribution, 
$\rho_p(\mathbf{r})$,
involving a single parameter, and after 
determining this parameter from the measured  value 
of $\langle r^2 \rangle _p$,
we deduce $\langle r \rangle _p$ from the assumed
$\rho_p(\mathbf{r})$. 
Table~\ref{tab:FS} gives $\langle r \rangle _p$
and $\langle r^2 \rangle _p$ calculated for three different 
functional forms of $\rho_p(\mathbf{r})$. 
The results for the exponential form,  
$\rho_p(\mathbf{r})\!=\!1/(8\pi  r^3_0){\rm e}^{-(r/r_0)}$,
are given in the fourth column;
the exponential form corresponds to a dipole-type 
proton form factor (in momentum space),
which reproduces very well the elastic electron-proton scattering data. 
We also present the results 
for two other commonly used forms for $\rho_p(\mathbf{r})$,  
the uniform distribution (second column),
and the Gaussian form (third column);
these results have been extracted from Ref.~\cite{Friar1979}. 
\begin{table}[h]
\caption{First and second moments of the proton charge distribution
calculated for various forms of $\rho_p(\mathbf{r})$
characterized by a single parameter.}

\vspace{4mm}
\begin{tabular}{c|c|c|c}
\hline\hline
       & Uniform & Gaussian & Exponential  
\\
$\rho_p(\mathbf{r})$ & $\frac{3}{4\pi R^3}\,\theta(R-r)$ &
$\left(\frac{1}{\sqrt{\pi}r_0}\right)^3\!{\rm e}^{-(r/r_0)^2}$ & 
$\frac{1}{8\pi  r^3_0}\,{\rm e}^{-(r/r_0)}$ 
\\[1ex]
\hline
$\langle r \rangle_p$ & $3R/4$ & $2r_0/\!\sqrt{\pi}$ & $3r_0$ 
 \\
$\langle r^2 \rangle_p$ & $3R^2\!/5$ & $3r^2_0/2$ & $12\,r^2_0$ 
\\[1ex]
\hline
$\langle r \rangle_p/\!\sqrt{\langle r^2 \rangle}_p$ & $\sqrt{15}/4=0.968$ &
$2\sqrt{2/3\pi}=0.931$ & $\sqrt{3}/2=0.866$ 
\\
\hline
\end{tabular}
\label{tab:FS}
\vspace{8mm}
\end{table}
The last row in table~\ref{tab:FS} shows
the ratio $\langle r_p \rangle/\!\sqrt{\langle r^2 \rangle}_p$
for each assumed form of $\rho_p(\mathbf{r})$. 
By taking the average of the results for these three cases,
we deduce
$\langle r \rangle_p=(0.916\pm0.051)\sqrt{\langle r^2 \rangle}_p$;
the ``error estimate" here has been obtained 
by interpreting the scatter of the results in table~\ref{tab:FS}
as a measure of uncertainty.
Then, with the use of the experimental value 
of the proton r.m.s. radius,
$\sqrt{\langle r^2 \rangle}_p=0.862$ fm~\cite{Simon1980}\footnote{ 
It is to be noted that a recent muonic hydrogen atom experiment
has questioned this value for the proton r.m.s. radius, 
see e.g., Ref.~\cite{Lorenz2012}
}, 
we find $\langle r \rangle _p = 0.790\pm 0.044$ fm.
Using this value in Eq.(\ref{eq:FS}) leads to 
$\delta\psi^{\rm FS}_{1s}({\mathbf 0}) 
\simeq -0.00275(1\pm0.056)$.
Correspondingly, the finite-proton-size correction
to the capture rate $\Gamma_S$ in Eq.(\ref{eq:C-rate}) 
is found to be  
$2\,\delta\psi^{\rm FS}_{1s}({\mathbf 0}) =-0.0055(1\pm0.06)$.
This result is essentially the same as that given in Eq.(8)
of Ref.~\cite{Sirlin2007}. 
Thus, the finite-proton-size correction is of the same order as   
the vacuum polarization correction shown in Table~\ref{tab:QED}. 

In addition to the two QED corrections discussed above,
we need to consider the ``standard" radiative corrections 
involving a weak-interaction vertex. 
It is to be noted that part of these corrections are already included 
in $G_F$, if one uses (as we do here)
the value of $G_F$ determined from the measured muon lifetime.
In the following, what we simply call the ``electroweak loop corrections" 
refer to those electroweak loop corrections that have not
been accounted for by the use of the $G_F$ 
derived from the muon lifetime.
%
We remark that, to the order in HB$\chi$PT under consideration,
the electroweak loop corrections are identical for $\mu p$ capture 
and neutron beta-decay.
We can therefore utilize the results obtained for neutron beta-decay 
in, e.g., Refs.~\cite{Ando2004,Sirlin1967}.
Since the muon velocity, $\beta$, in the initial $\mu p$-atomic state
is essentially zero, we can take the limit of  $\beta\!\to\!0$
in the previous evaluations of the radiative corrections 
to the neutron beta-decay rate~\cite{Sirlin1967,Ando2004}, 
(In applying the results obtained for neutron $\beta$-decay
to the $\mu p$ capture case,
we must drop the bremsstrahlung contributions,
since both the initial and final states in $\mu p$ capture
contain only charge-neutral particles.)
Then the electroweak radiative loop correction 
to the $\mu p$ capture rate is obtained as 
\begin{eqnarray}
\Gamma_S^{(0)}\to\Gamma_S^{(0)}(1+RC_{EW})\,,
\label{eq:EWa}
\end{eqnarray}

\vspace{-5mm}
\noindent
with
\begin{eqnarray}
RC_{\rm EW}\,=\frac{\alpha}{2\pi}\!\left\{ 
\tilde{e}_V^R (m_N) + 
3\ln\!\left[\frac{m_N}{m_\mu}\right]-\frac{27}{4}
\right\}
\label{eq:EW}
\end{eqnarray} 

\vspace{2mm}
\noindent
In this expression the electroweak LEC, $\tilde{e}^R_V(m_N)$,
subsumes short-distance physics not probed in the low-energy muon capture reaction. 
The value of this LEC at the scale, $\lambda=\!m_N$, 
has been determined in Refs.~\cite{Ando2004,Udit2011} 
by comparing with the 
expressions for the short-distance radiative corrections derived by 
Sirlin and Marciano~\cite{Sirlin1986,Sirlin1974} for the electroweak processes.
The result is  $\tilde{e}^R_V(m_N)= 19.5$. 
In the next section we discuss 
the numerical consequences of our evaluation of the ${\mathcal O}(\alpha)$
radiative and finite proton-size corrections discussed in this section.

\section{Numerical Results for the Capture Rates, $\Gamma_0$ and $\Gamma_1$}

As explained earlier, $\Gamma_0^{(0)}$ ($\Gamma_1^{(0)}$) denotes
the hyperfine-singlet (hyperfine-triplet) capture rate
calculated without including radiative corrections;
{\it viz.,} $\Gamma_0^{(0)}$ and $\Gamma_1^{(0)}$
are obtained by identifying
$\Psi_{\mu p}({\mathbf 0})$ in Eq.(\ref{eq:C-rate})
with $\Phi_{1s}({\mathbf  0})$.
Using the inputs listed in Table~\ref{tab:ff}, we obtain

\vspace{-12mm}
\begin{eqnarray}
\Gamma_0^{(0)} = 693\,{\rm s}^{-1}\,\,\,\,\,{\rm and}\,\,\,\,\,\,
\Gamma_1^{(0)} = 12.0\,{\rm s}^{-1}\,,
\end{eqnarray} 
corresponding to the use of  
$g_A \!= \!1.266$ and $g_{\pi NN}\!=\!13.40$. 
The inclusion of the radiative corrections discussed
in Section III modifies $\Gamma_S^{(0)}$ ($S=0,\,1$)
to $\Gamma_S$ as
\begin{equation}
\Gamma_S=\Gamma_S^{(0)}(1+RC_{\rm QED}+RC_{\rm EW})
\;\:;\;\; S=0,\,1.
\end{equation}
Here $RC_{\rm QED}$ represents the corrections arising from
the change in the atomic $\mu p$ wave function 
due to the vacuum-polarization 
and finite-proton-size effects,
while, as explained earlier, $RC_{\rm EW}$ is 
the electroweak radiative correction:
\begin{eqnarray}
RC_{\rm QED}&=&2\delta\psi^{\rm VP}_{1s}({\mathbf 0})+
2\delta\psi^{\rm FS}_{1s}({\mathbf 0})
\label{eq:RC1}
\\
RC_{\rm EW}\,&=&\frac{\alpha}{2\pi}\!\left\{ 
\tilde{e}_V^R (m_N) + 
3\ln\!\left[\frac{m_N}{m_\mu}\right]-\frac{27}{4}
\right\}\, .
\label{eq:RC2}
\end{eqnarray}

\vspace{1mm}
\noindent
We remark that, since the last two terms in Eq.(\ref{eq:RC2})
almost cancel each other, 
$RC_{\rm EW}$ has a pronounced dependence on the LEC, 
$\tilde{e}_V^R (m_N)$, 
which characterizes the short-distance processes.

\vspace{5mm}
The numerical consequences of including 
the radiative corrections are displayed in Table~\ref{tab:final},
where $\Gamma_S$ ($S\!=\!0,\,1$) are shown 
along with $\Gamma_S^{(0)}$ and the changes due to the
individual contributions of $RC_{\rm QED}$ and $RC_{\rm EW}$.
Again, these results have been obtained with the use of
$g_A \!= \!1.266$ and $g_{\pi NN}\!=\!13.40$.
Table~\ref{tab:final} demonstrates that the largest radiative 
correction to the OMC rate comes from  $RC_{\rm EW}$, 
in conformity with the results reported in Ref.~\cite{Sirlin2007}.
In particular, for the hyperfine-singlet OMC rate, which is of our main concern,
$RC_{\rm EW}$ changes $\Gamma_0^{(0)}$ by $\sim\!2$ \%.
 \begin{table}[h]
\label{tab:final}
\caption{The hyperfine-singlet and -triplet OMC rates, $\Gamma_0$ and $\Gamma_1$
(in units of $s^{-1}$), calculated with and without radiative corrections (the
proton-finite-size effect is included as part of $RC_{\rm QED}$) 
corresponding to $g_{\pi NN}\!=\!13.40$ and $g_A\!=\!1.266$.}
        
\vspace{5mm}
\begin{tabular}{c|c|c|c}
\hline\hline
$\Gamma_0^{(0)}$ \,&\, $\Gamma_0^{(0)}\!(1\!+\!RC_{\rm QED})$  \,&\,
$\Gamma_0^{(0)}\!(1\!+\!RC_{\rm EW})$ \,&\,
$\Gamma_0=\Gamma_0^{(0)}\!(1\!+\!RC_{\rm QED}\!+\!RC_{\rm EW})$ 
\\[1ex]
\hline
 692.9 \,& 691.7  & 708.4 & 707.2  \\
\hline\hline
$\Gamma_1^{(0)}$  \,&\, $\Gamma_1^{(0)}\!(1\!+\! RC_{\rm QED})$  \,&\,
$\Gamma_1^{(0)}\!(1\!+\! RC_{\rm EW})$ \,&\,
$\Gamma_1=\Gamma_1^{(0)}\!(1\!+\! RC_{\rm QED}\!+\!RC_{\rm EW})$
\\[1ex]
\hline
 12.0 \,& 11.9  & 12.2 & 12.2 \\
\hline
\end{tabular}
\vspace{10mm}
\end{table}

\vspace{2mm}
\section{Discussion and Conclusions}
In the previous section we have presented our numerical results 
obtained with the use of representative values 
for the relevant input parameters.
We now discuss to what extent the uncertainties 
in these input parameters affect the calculated values 
of  the $\mu p$ capture rates, $\Gamma_S$ ($S\!=\!0,1$).
We shall chiefly concentrate on the hyperfine-singlet rate
$\Gamma_0$,
a quantity of primary concern for most $\mu p$ capture experiments.

As mentioned, the LEC, $\tilde{B}_2$, is determined from the G-T discrepancy
[see Eq.(\ref{eq:B2-GT})], and the fact that the current precision of the values 
of $g_A$ and $g_{\pi NN}$ is somewhat limited leads to rather significant 
uncertainty in $\tilde{B}_2$. 
The results in Table~\ref{tab:final}
were obtained for ${\tilde{B}}_2= -1.761$, 
which corresponds to $g_{\pi NN}\!=\!13.40$ and $g_A\!=\!1.266$
taken from PDG2002~\cite{PDG2002}.
If we adopt  $g_{\pi NN}\!=\!13.05$ and $g_A\!=\!1.270$
(values given in PDG2012~\cite{RPP2012}),
then we obtain ${\tilde{B}}_2= -0.498$
and, correspondingly, 
$\Gamma_0^{(0)} = 699\,s^{-1}$  and $\Gamma_1^{(0)} = 11.7$~$s^{-1}$. 
Thus, the uncertainty in $\tilde{B}_2$ changes $\Gamma_0^{(0)}$ by 
7 $s^{-1}$ (about 1\% increase), and $\Gamma_1^{(0)}$ by 0.2~$s^{-1}$ (about 2 \% decrease). 
If we take into account (in the last column in Table~\ref{tab:final})
the mentioned variation in $\Gamma_0^{(0)}$,
the corresponding change in $\Gamma_0$ ranges 
from  707.2~$s^{-1}$ to 713.7~$s^{-1}$; thus
\begin{equation}
\Gamma_0=710.4\!\times\!(1\pm0.005)\,s^{-1},
\label{eq:G0a}
\end{equation}
where the relative error was deduced from the 1 \% difference
between the above-quoted two values of $\Gamma_0^{(0)}$.
  
We next consider the uncertainty in the proton axial radius,
$(\langle r_A^2 \rangle)^{1/2}$, discussed in Section II.
The results shown in Table~\ref{tab:final}
were obtained for $(\langle r_A^2 \rangle)^{1/2}= 0.65$ fm.
If we instead use  $(\langle r_A^2 \rangle)^{1/2} = 0.57$ fm,
corresponding to $m_A=1200$ MeV,
we find $\Gamma_0^{(0)} = 695.7  s^{-1}$, 
an increase of 2.9 $s^{-1}$ 
(or $\sim$ 0.5\%). 
Again, if we consider (in the last column in Table~\ref{tab:final})
the scatter in the value of $\Gamma_0^{(0)}$, then the corresponding 
change in $\Gamma_0$ ranges from 707.2~$s^{-1}$ to 710.1~$s^{-1}$, i.e.,
\begin{equation}
\Gamma_0=708.6\!\times\!(1\pm0.0025)\,s^{-1}\,
\label{eq:G0b}
\end{equation}
where the relative error was deduced from the 0.5 \% variation
in $\Gamma_0^{(0)}\!$. Taking the average of the values in 
Eqs.(\ref{eq:G0a}) and (\ref{eq:G0b}),
we arrive at
\begin{equation}
\Gamma_0=710\!\times\!(1\pm0.006)\,s^{-1}\,,
\label{eq:GammaSummary}
\end{equation}
where the error has been deduced from the quadratic sum
of the errors in Eqs.(\ref{eq:G0a}) and (\ref{eq:G0b}).

In connection with Table~\ref{tab:final} we have pointed out 
that, of all the corrections of $\cal{O}(\alpha)$,
the electroweak loop correction, $RC_{\rm EW}$,
is largest; it increases $\Gamma_0^{(0)}$ by 
as much as $\sim$2 \%.
So, if $RC_{\rm EW}$ is not evaluated with sufficient accuracy,
the theoretical error in $\Gamma_0$ can be larger 
than indicated by Eq.(\ref{eq:GammaSummary}).
As already mentioned, $RC_{\rm EW}$ is a sensitive function 
of the LEC, $\tilde{e}_V^R (M_N)$, as the last two terms in Eq.(\ref{eq:RC2})  
nearly cancel each other.
In the present work, following Ref.\cite{Ando2004}, 
we have determined $\tilde{e}_V^R (M_N)$
by comparing our HB$\chi$PT results with 
those obtained in the S-M method~\cite{Sirlin1986,Sirlin1974}.
Since this method is generally considered to be highly reliable,
we believe that $\tilde{e}_V^R (M_N)$ is known with sufficient accuracy 
to make the uncertainty in $\Gamma_0$
related to $RC_{\rm EW}$ much smaller than 0.6 \%,
the error arising from the other sources [see Eq.(\ref{eq:GammaSummary})].
We remark that the same LEC, $\tilde{e}_V^R (m_N)$, 
also appears in neutron beta decay~\cite{Ando2004}
and the inverse beta decay process,
$\bar{\nu}_e \!+\!p\to e^+ \!+\!n$~\cite{Udit2011}. 
It is therefore, in principle, possible to use either the neutron
$\beta$-decay or  the $\mu p$ capture to control $\tilde{e}_V^R (m_N)$
and make predictions for the other processes involving the same LEC.
This would allow us to deduce $\tilde{e}_V^R (m_N)$
without using the result of the S-M method.

In conclusion, the present HB$\chi$PT calculation 
of the hyperfine-singlet $\mu p$ capture rate $\Gamma_0$,
including radiative and proton finite-size corrections 
of ${\mathcal O}(\alpha)$, gives
\begin{equation}
\Gamma_0=710 \pm 5\,s^{-1}\,.
\label{eq:GammaFinal}
\end{equation}
This is in excellent agreement with the experimental value
quoted in Eq.(\ref{eq:Gammaexp}).
The 0.6~\%  theoretical error in Eq.(\ref{eq:GammaFinal})
is dominated by the uncertainties in the input values
of $g_A$ and $g_{\pi NN}$ that enter into the G-T discrepancy.

\vspace{5mm}
\noindent
{\bf Acknowledgements}\\
This work is supported in part by grants from the National Science Foundation,
PHY-0758114 and PHY-1068305.



\begin{thebibliography}{99}  




\bibitem{MuCap2007} 
V.A. Andreev {\it et al.}
(MuCap Collaboration), Phys. Rev. Lett. {\bf 99}, 032002 (2007); \\
V.A. Andreev {\it et al.}
(MuCap Collaboration), Phys. Rev. Lett. {\bf 110}, 012504 (2013), 
arXiv:1210.6545 [nucl-ex].

\bibitem{Primakoff1975} 
H. Primakoff, in {\it Nuclear and Particle Physics at Intermediate Energies}
(Plenum, New York, 1975). 

\bibitem{Fearing2003}
T. Gorringe and H.W. Fearing, Rev. Mod. Phys. {\bf 76}, 31 (2003). 

\bibitem{Kammel2010} 
P. Kammel and K. Kubodera, Annu. Rev. Nucl. Part.  Sci. {\bf 60}, 327 (2010). 

\bibitem{bkm94}
V. Bernard, N. Kaiser and U.-G. Mei\ss ner, 
Phys. Rev. D, {\bf 50}, 6899 (1994). 

\bibitem{Adler1966}
S.L. Adler and Y. Dothan, Phys. Rev. {\bf 151}, 1267 (1966).

\bibitem{Wolfenstein1970}
L. Wolfenstein, in {\it High-Energy Physics and Nuclear Structure}, 
ed. S. Devons 
(Plenum, New York, 1970), p. 661. 


\bibitem{Kaiser2003}
N. Kaiser, Phys. Rev. C, {\bf 67 }, 027002 (2003).


\bibitem{Sirlin2007} 
A. Czarnecki, W.J. Marciano and A. Sirlin, 
Phys. Rev. Lett. {\bf 99}, 032003 (2007)

\bibitem{Sirlin1974}
A. Sirlin, Nucl. Phys. B{\bf 71}, 29 (1974);
A. Sirlin, Nucl. Phys. B{\bf 100}, 291 (1975); 
A. Sirlin, arXiv:hep-ph/0309187 (2003)

\bibitem{Sirlin1986}
W.J. Marciano and A. Sirlin, Phys. Rev. Lett. {\bf 56}, 22 (1986).

\bibitem{bkm95} V. Bernard, N. Kaiser and U.-G. Mei\ss ner, 
Int. J. Mod. Phys. E {\bf 4},  193  (1995)

\bibitem{Bernard2009} 
V. Bernard, Prog. Nucl. Part. Phys. {\bf 60}, 82 (2008);
arXiv:0706.0312[hep-ph].  

\bibitem{Scherer} 
S. Scherer, Prog. Nucl. Part. Phys. {\bf 64}, 1 (2010); 
arXiv:0908.3425[hep-ph].

\bibitem{Ando2004}
S. Ando, H.W. Fearing, V. Gudkov, K. Kubodera, F. Myhrer, 
S. Nakamura and T. Sato,
Phys. Lett. B {\bf 595}, 250 (2004).

\bibitem{Udit2011} U. Raha, F. Myhrer and K. Kubodera, 
Phys. Rev. C {\bf 85}, 045502 (2012);  
arXiv:1112.2007[hep-ph];
U. Raha, F. Myhrer and K. Kubodera, 
arXiv:1207.4306 [nucl-th].

\bibitem{Ando2000} 
S. Ando, F. Myhrer and K. Kubodera, 
Phys. Rev. C, {\bf 63}, 015203 (2000). 

\bibitem{bhm2001}
V. Bernard, T.R. Hemmert and U.-G. Mei\ss ner, 
Nucl. Phys. A, {\bf 686}, 290 (2001). 

\bibitem{Bernard2002}
V. Bernard, L. Elouadrhiri and U.-G. Mei\ss ner, 
J. Phys. G: Nucl. Part. Phys. {\bf 28}, R1 (2002). 

\bibitem{Bernard1998}
V. Bernard, H.W. Fearing, T.R. Hemmert  and U.-G. Mei\ss ner, 
Nucl. Phys.  {\bf A635}, 121 (1998); Nucl. Phys.  {\bf A642}, 563 (1998). 

\bibitem{Fearing1997} 
H.W. Fearing, R. Lewis, N. Mobed and S. Scherer,
Phys. Rev. D {\bf 56}, 1783 (1997);
H.W. Fearing, R. Lewis, N. Mobed and S. Scherer,
Nucl. Phys. {\bf A631}, 735c (1998).

\bibitem{PDG2002}
K. Higawara {\it et al.} 
(Particle Data Group), Phys. Rev. D, {\bf 66}, 010001 (2002).

\bibitem{RPP2012}
J. Beringer {\it et al.} (Particle Data Group), Phys. Rev. D {\bf 86}, 010001 (2012).


\bibitem{Eiras2000}
D. Eiras and J. Soto, Phys. Lett. B {\bf 491}, 101 (2000). 

\bibitem{Friar1979}
J.L. Friar, Ann. Phys. {\bf 122}, 151 (1979).

\bibitem{Goldman1972}
M. R. Goldman, Nucl. Phys. B, {\bf 49}, 621 (1972). 

\bibitem{Simon1980} 
G.G Simon, Ch. Schmitt, F. Borkowski and V.H. Walter,
Nucl. Phys. A, {\bf 333}, 381 (1980). 

\bibitem{Lorenz2012}
I.T. Lorenz, H.-W. Hammer and U.-G. Mei\ss ner, arXiv:1205.6628 [hep-ph]; 
R. Pohl, R. Gilman, G.A. Miller, K. Pachucki,  
arXiv:1301.0905 [physics.atom-ph]. 

\bibitem{Sirlin1967}
A. Sirlin, Phys. Rev. {\bf 164}, 1767 (1967).

\end{thebibliography}
\end{document}